\documentclass[a4paper]{article}
\usepackage[german, english]{babel}
\usepackage{a4wide}
\usepackage{amsmath}
\usepackage{amssymb}
\usepackage{ifthen}
\usepackage{epsfig}
\newcounter{fig}   

\newcommand{\vphi}{\varphi}

\newcommand{\sqt}{\sqrt{3}}

\begin{document}

\title{\bf Platonic Sphalerons\\ in the Presence of a Dilaton Field}
\vspace{1.5truecm}
\author{
{\bf Burkhard Kleihaus, Jutta Kunz and Kari Myklevoll}\\
Institut f\"ur  Physik, Universit\"at Oldenburg, Postfach 2503\\
D-26111 Oldenburg, Germany}

\vspace{1.5truecm}

\date{\today}

\maketitle
\vspace{1.0truecm}

\begin{abstract}
We construct sphaleron solutions with discrete symmetries
in Yang-Mills-Higgs theory coupled to a dilaton.
Related to rational maps of degree $N$, 
these platonic sphalerons can be assigned a Chern-Simons number 
$Q=N/2$.
We present sphaleron solutions with degree $N=1-4$,
possessing spherical, axial, tetrahedral and cubic symmetry.
For all these sphalerons two branches of solutions exist,
which bifurcate at a maximal value of the dilaton coupling constant.
\end{abstract}
%\vfill\eject

\section{Introduction}

The non-trivial topology of the configuration space of Weinberg-Salam theory
gives rise to unstable classical solutions, sphalerons \cite{km,bi},
associated with baryon number violating processes \cite{review}.
Besides the well-known single sphalerons \cite{km}, 
Weinberg-Salam theory allows for
sphaleron-antisphaleron systems \cite{kl},
as well as for multisphalerons \cite{kksph,kkm}.
Multisphalerons with axial symmetry have long been known \cite{kksph},
but only recently multisphalerons with no rotational symmetry at all
have been found \cite{kkm}.
The symmetries of these sphalerons are only discrete,
and can be identified with the symmetries of platonic solids or crystals.
We therefore refer to them as platonic sphalerons.
Related to certain rational maps of degree $N$ \cite{ratmap},
they have many properties in common with
platonic monopoles and Skyrmions \cite{plato1,plato2}.

When gravity is coupled to the bosonic sector of Weinberg-Salam theory,
i.e.~to Yang-Mills-Higgs (YMH) theory with a Higgs doublet,
the flat space sphaleron changes smoothly, and a branch
of gravitating sphalerons arises \cite{greene,vg,yves}.
This branch bifurcates at a maximal value of the gravitational
coupling constant with a second branch, higher in energy.
In the limit of vanishing coupling constant,
this second branch ends at the (lowest) Bartnik-McKinnon solution \cite{bm}
of Einstein-Yang-Mills (EYM) theory.

The coupling of gravity to YM or YMH theory
has a very similar effect 
concerning the existence of classical solutions as
the coupling to a scalar dilaton \cite{lav,gmono,forgacs}.
For monopoles, arising in YMH theory with Higgs triplet,
even the maximal values of the coupling constants
for gravity and the dilaton are very close \cite{gmono,forgacs}.
This suggests, that the influence of gravity on
the new platonic sphalerons may be to some extend mimicked
by the presence of a scalar dilaton.

With the aim in mind to obtain self-gravitating objects,
which possess only discrete, platonic symmetries,
we therefore here present as a first step in this direction
the study of dilatonic sphalerons with only discrete symmetries,
corresponding to tetrahedral ($N=3$) and cubic ($N=4$) sphalerons.
They emerge from the platonic sphalerons of Weinberg-Salam theory
in the limit of vanishing dilaton coupling.
Forming two branches, which bifurcate at a maximal value
of the dilaton coupling constant, they indicate the
existence of platonic solutions also in Yang-Mills-dilaton (YMD) theory.
For comparison we also present spherically symmetric ($N=1$)
and axially symmetric $(N=2-4$) sphalerons.

We review Yang-Mills-Higgs-dilaton (YMHD) theory in section 2. 
We present the Ans\"atze and the boundary conditions for axially symmetric
and platonic sphalerons in section 3,
and discuss our numerical results in section 4.

\section{Yang-Mills-Higgs-Dilaton Theory}

We consider Yang-Mills-Higgs-dilaton (YMHD) theory with Lagrangian
\begin{equation}
{\cal L} = -\frac{1}{2} \partial_\mu \phi \partial^\mu \phi
-\frac{1}{2} e^{2\kappa \phi} {\rm Tr} (F_{\mu\nu} F^{\mu\nu})
- (D_\mu \Phi)^\dagger (D^\mu \Phi) 
- \lambda e^{-2\kappa \phi} (\Phi^\dagger\Phi - \frac{v^2}{2} )^2
\  
\label{lag1}
\ , \end{equation}
SU(2) field strength tensor
\begin{equation}
F_{\mu\nu}=\partial_\mu V_\nu-\partial_\nu V_\mu
            + i g [V_\mu , V_\nu ]
\ , \end{equation}
SU(2) gauge potential $V_\mu = V_\mu^a \tau_a/2$,
covariant derivative of the Higgs doublet $\Phi$
\begin{equation}
D_{\mu} \Phi = \partial_{\mu}\Phi+i g V_\mu  \Phi
\ , \end{equation}
and dilaton field $\phi$,
where $g$ and $\kappa$ denote the gauge and dilaton coupling constants,
respectively,
$\lambda$ denotes the strength of the Higgs self-interaction, and
$v$ the vacuum expectation value of the Higgs field.

The Lagrangian (\ref{lag1}) is invariant under local SU(2)
gauge transformations $U$,
\begin{eqnarray}
V_\mu &\longrightarrow & U V_\mu U^\dagger
+ \frac{i}{g} \partial_\mu U  U^\dagger \ ,
\nonumber\\
\Phi  &\longrightarrow & U \Phi \ .
\nonumber
\end{eqnarray}
The gauge symmetry is spontaneously broken 
due to the non-vanishing vacuum expectation
value of the Higgs field
\begin{equation}
    \langle \Phi \rangle = \frac{v}{\sqrt2}
    \left( \begin{array}{c} 0\\1  \end{array} \right)   
\ , \end{equation}
leading to the boson masses
\begin{equation}
    M_W = M_Z = \frac{1}{2} g v \ , \ \ \ \ 
    M_H = v \sqrt{2 \lambda} \ . 
\end{equation}
In the limit of vanishing dilaton coupling constant,
the model corresponds to the bosonic sector of the
Weinberg-Salam theory for vanishing Weinberg angle.

In the following we consider only static finite energy solutions, with
$V_0=0$, $V_i = V_i(\vec{r}\,)$, $i=1,2,3$, $\Phi = \Phi(\vec{r}\,)$, 
$\phi = \phi(\vec{r}\,)$.
The energy of such solutions is given by 
\begin{equation}
E = \int\left( \frac{1}{2} \partial_i \phi \partial^i \phi
+\frac{1}{2} e^{2\kappa \phi} {\rm Tr} (F_{ij} F^{ij})
+ (D_i \Phi)^\dagger (D^i \Phi) 
+ \lambda e^{-2\kappa \phi} (\Phi^\dagger \Phi - \frac{v^2}{2})^2  
\right) d^3r \ .
\label{energy}
\end{equation}
The energy $E$ and
the dilaton charge $D$, where
\begin{equation}
D = \frac{1}{4\pi} \int_{S_2} \vec{\nabla}\phi \cdot d\vec{S} \ ,
\end{equation}
are related by 
\begin{equation}
4 \pi D = \kappa E \ ,
\label{EDrel}
\end{equation}
as can be seen by integrating the dilaton equation and using the identity
\begin{equation}
0=\int\left( -\frac{1}{2} \partial_i \phi \partial^i \phi
+\frac{1}{2} e^{2\kappa \phi} {\rm Tr} (F_{ij} F^{ij})
+ (D_i \Phi)^\dagger (D^i \Phi) 
+ 3 \lambda e^{-2\kappa \phi} (\Phi^\dagger \Phi - \frac{v^2}{2} )^2 
\right) d^3r
\ , \label{derrick} 
\end{equation}
which follows from a Derrick like argument
for solutions of the field equations.

\section{Sphaleron Solutions}

Let us consider the Ansatz and boundary conditions first for the 
axially symmetric sphaleron solutions, 
and then for the sphaleron
solutions with platonic symmetries.

\subsection{Axially Symmetric Sphalerons}

The Ansatz for the axially symmetric YMHD sphalerons 
corresponds to the Ansatz employed in the Weinberg-Salam theory
(at vanishing Weinberg angle) \cite{kksph},
\begin{equation}
V_i dx^i = \left(\frac{H_1}{r} dr + (1-H_2) d\theta\right)
           \frac{\tau^{(n)}_\vphi}{2g}
          -n\sin\theta\left(H_3 \frac{\tau^{(n)}_r}{2g} 
	  + (1-H_4)\frac{\tau^{(n)}_\theta}{2g}\right) d\vphi
	   \ , \ \ \ V_0=0 \ ,
\label{a_axsym}
\end{equation}	  
and
\begin{equation}
\Phi = i( \Phi_1 \tau^{(n)}_r + \Phi_2 \tau^{(n)}_\theta )\frac{v}{\sqrt2}
    \left( \begin{array}{c} 0\\1  \end{array} \right) \ ,
\end{equation}
supplemented by the dilaton function $\phi$,
where
\begin{eqnarray}	  
\tau^{(n)}_r & = & \sin\theta (\cos n\vphi \tau_x + \sin n\vphi \tau_y) 
           + \cos\theta \tau_z \ , \ \ 
\nonumber \\	   
\tau^{(n)}_\theta & = & \cos\theta (\cos n\vphi \tau_x + \sin n\vphi \tau_y) 
           - \sin\theta \tau_z \ , \ \ 
\nonumber \\	   
\tau^{(n)}_\vphi & = & (-\sin n\vphi \tau_x + \cos n\vphi \tau_y) 
\ , \ \ \nonumber 
\end{eqnarray}	  
and $\tau_x$, $\tau_y$ and $\tau_z$ denote the Pauli matrices.
The integer $n$ is related to the Chern-Simons charge of the
sphalerons,
$Q=N/2$, where $n=N$ \cite{kksph}.
For $n=1$ and $\kappa=0$ the Ansatz yields the 
spherically symmetric Klinkhamer-Manton sphaleron \cite{km}.
For $n>1$,
the functions $H_1$--$H_4$, $\Phi_1$, $\Phi_2$ and 
$\phi$ depend on $r$ and $\theta$, only. 
With this Ansatz the full set of field equations reduces to a system 
of seven coupled partial differential equations in the independent variables 
$r$ and $\theta$. A residual U(1) gauge degree of freedom is 
fixed by the condition $r\partial_r H_1 - \partial_\theta H_2=0$ \cite{kksph}.

Regularity at the origin and on the $z$-axis requires 
\begin{equation}
H_1=H_3=\Phi_1=\Phi_2=0 \ , \ H_2=H_4=1 \ , \ \partial_r \phi =0 \ , \ \ \ \ 
{\rm at} \ \ r=0 \ ,
\end{equation}
respectively
\begin{equation}
H_1=H_3=\Phi_2=0 \ , \ 
\partial_\theta H_2=\partial_\theta H_4
=\partial_\theta \Phi_1=\partial_\theta \phi=0 \ , \ \ \ \  
{\rm at} \ \ \theta=0 \, ,  \pi \ ,
\end{equation}
while the condition of finite energy implies
\begin{equation}
H_1=H_3=\Phi_2=0 \ , \ H_2=H_4=-1 \ , \Phi_1 = 1 \  , \ \phi =\phi_\infty \ ,  
\ \ \ \ {\rm as} \ \ r \to \infty \ .
\end{equation}

\subsection{Platonic Sphalerons}

To obtain YMHD solutions with discrete symmetry
we make use of rational maps,
i.e.~holomorphic functions from $S^2\mapsto S^2$ \cite{ratmap}.
Treating each $S^2$ as a Riemann sphere, the first having coordinate 
$\xi$,
a rational map of degree $N$ is a function $R:S^2\mapsto S^2$ where
\begin{equation}
R(\xi)=\frac{p(\xi)}{q(\xi)} 
\ , \label{rat} \end{equation}
and $p$ and $q$ are polynomials of degree at most $N$, where at least
one of $p$ and $q$ must have degree precisely $N$, and $p$ and $q$
must have no common factors \cite{ratmap}.

We recall that via stereographic projection, the complex coordinate $\xi$
on a sphere can be identified with conventional polar coordinates by
$\xi=\tan(\theta/2)e^{i\varphi}$ \cite{ratmap}.
Thus the point $\xi$ corresponds to the unit vector
\begin{equation}
\vec {n}_\xi=\frac{1}{1+\vert \xi \vert^2}
(2\Re(\xi), 2\Im(\xi),1-\vert \xi \vert^2)
\ , \label{unit1} \end{equation}
and the value of the rational map $R(\xi)$ 
is associated with the unit vector
\begin{equation}
\vec {n}_R=\frac{1}{1+\vert R \vert^2}
(2\Re(R), 2\Im(R),1-\vert R\vert^2) \ .
\label{unit2}
\end{equation}

We here consider platonic YMHD solutions obtained from maps $R_N$,
\begin{equation}
R_3(\xi)=\frac{\sqrt{3}a\xi^2-1}{\xi(\xi^2-\sqrt{3}a)} \ , \ \ a=\pm i \ , \
\label{map1} \end{equation}
\begin{equation}
R_4(\xi)=c\frac{\xi^4+2\sqrt{3}i\xi^2+1}{\xi^4-2\sqrt{3}i\xi^2+1} \ , \ \ c=1 \ . \
\label{map2} \end{equation}
Note, that the choice $a=0$ in (\ref{map1}),
yields the axially symmetric sphalerons
for $N=3$ in a different gauge,
while the axially symmetric sphalerons for $N=4$ are obtained from 
$R_4(\xi)= \xi^4$.

Parametrizing the Higgs field as 
\begin{equation}
\Phi = (\Phi_0 1\hspace{-0.28cm}\perp + i \Phi_a \tau_a)\frac{v}{\sqrt2}
    \left( \begin{array}{c} 0\\1  \end{array} \right) \ ,
\end{equation}
we impose at infinity the boundary conditions
\begin{equation}
\Phi_0 = 0 \ , \ \ \ 
\Phi_a  \tau_a=   \vec {n}_R \cdot {\vec \tau} =: \tau_R\ . 
\label{bcHiggs} 
\end{equation}
The boundary conditions for the gauge field are then obtained from 
the requirement $D_i \Phi =0$ at infinity, yielding
\begin{equation}
V_i = \frac{i}{g} (\partial_i \tau_R )\tau_R
\ , 
\label{bcA} 
\end{equation}
i.~e.~the gauge field tends to a pure gauge at infinity, 
$V_i = \frac{i}{g} (\partial_i U_\infty) U_\infty^\dagger $, 
with $U_\infty=i \tau_R$.
For the dilaton field we require that it vanishes at infinity,
$\phi_{\infty}=0$,
since any finite value of the dilaton field at infinity
can always be transformed to zero via
$\phi \rightarrow \phi - \phi_\infty$,
$r \rightarrow r e^{-\kappa \phi_\infty} $.

Subject to these boundary conditions,
and the gauge condition
\begin{equation}
\partial_i V^i =0
\ , \label{gaugecond} \end{equation}
one can then solve the general set of field equations, involving
the dilaton function $\phi(x,y,z)$,
3 functions $\Phi_a(x,y,z)$ for the Higgs field, 
and 9 functions $V_i^a(x,y,z)$ for the gauge field,
and $V_0^a=\Phi_0=0$.

\section{\bf Numerical results}

For convenience
we rescale the coordinates $\vec{r} \to \vec{r}/g v$,
the gauge potential $V_i \to v V_i$, the dilaton field $\phi \to \phi/\kappa$
and the coupling constants $\lambda \to g^2 \beta^2/8$ and
$\kappa \to \alpha/v$. This leaves only the dimensionless
parameters $\beta = M_{\rm H}/M_{\rm W}$ and $\alpha = 2 M_{\rm W} \kappa/g$.
We also rescale the dilaton charge $D \to D/\kappa g v$ and
the energy $E \to E 4 \pi v /g$. In terms of the dimensionless quantities,
the dilaton charge--energy relation (\ref{EDrel}) becomes
\begin{equation}
D = \alpha^2 E
\label{EDreld}
\end{equation}

The numerical solutions are constructed with help of the 
software package FIDISOL \cite{fidisol} based on the Newton-Raphson 
algorithm. Typical grids contain $70\times 30$ points for 
the axially symmetric solutions and $50 \times 25 \times 25$ points
for the platonic solutions.
The estimated relative errors are approximately $\approx 0.01$\% 
for the axially symmetric sphalerons, and 
$\approx 1$\% 
for the $N=3$ and $\approx 0.1$\% for the $N=4$ platonic sphalerons.

We consider sphaleron solutions 
with spherical symmetry ($N=1$), axial symmetry ($N=2,3,4$),
tetrahedral symmetry ($N=3$) and cubic symmetry ($N=4$).
We construct these solutions for the
Higgs masses $M_{\rm H}=0$ and $M_{\rm H}=M_{\rm W}$,
and study their dependence on the dilaton coupling constant $\alpha$.
The solutions are obtained in spherical coordinates $r$, $\theta$, $\vphi$.
To map the infinite range of the radial variable $r$ to the finite 
interval $[0,1]$ we introduce the compactified variable $\bar{r}=r/(1+r)$.
In the following we discuss successively the spherically symmetric 
($N=1$), axially symmetric ($N>1$)
and platonic ($N=3$, $N=4$) sphalerons.

\subsection{Spherically symmetric sphalerons}

Spherically symmetric dilatonic sphalerons have been considered
before \cite{Manka}.
When the dilaton coupling constant $\alpha$ is increased from zero,
a branch of dilatonic sphalerons emerges
from the spherically symmetric sphaleron of the Weinberg-Salam theory.
This branch extends up to a maximal value $\alpha_{\rm max}$  
of the coupling parameter $\alpha$, 
where a bifurcation with a second branch of solutions occurs.
The second branch then again extends backwards to $\alpha=0$.
On the first branch the energy decreases with increasing $\alpha$, 
whereas on the second branch it increases with decreasing $\alpha$ 
and diverges as $\alpha$ tends to zero.
We refer to these two branches as the lower and upper branch, 
respectively.
At the same time the value of the dilaton function at the origin
decreases continuously along both branches.
The energy of the spherically symmetric sphaleron
is shown as a function of $\alpha$ in Fig.~1a for 
$M_{\rm H}=0$ and $M_{\rm H}=M_{\rm W}$,
while the value of the dilaton function at the origin
is shown in Fig.~1b.
We note that the value of $\alpha_{\rm max}$ decreases with increasing 
Higgs mass $M_{\rm H}$.

\parbox{\textwidth}{
\centerline{
(a)\mbox{\epsfysize=5.0cm \epsffile{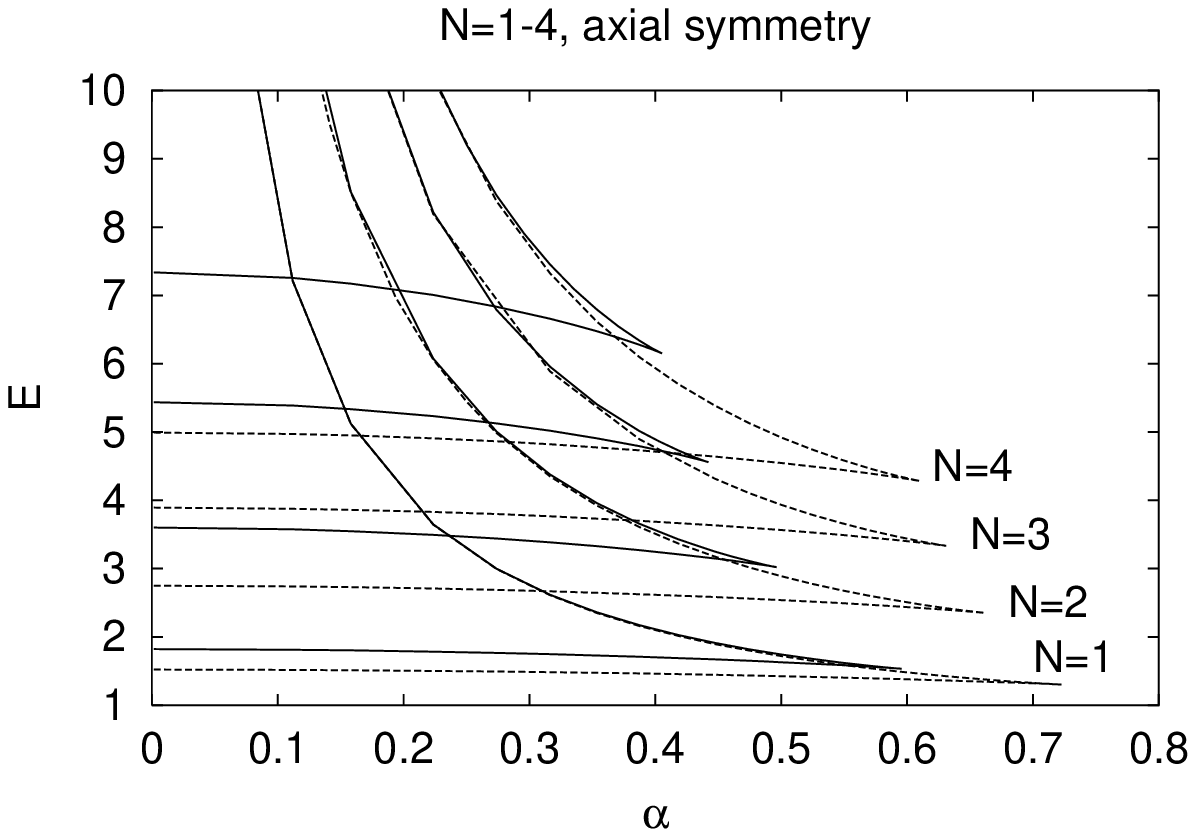} } 
(b)\mbox{\epsfysize=5.0cm \epsffile{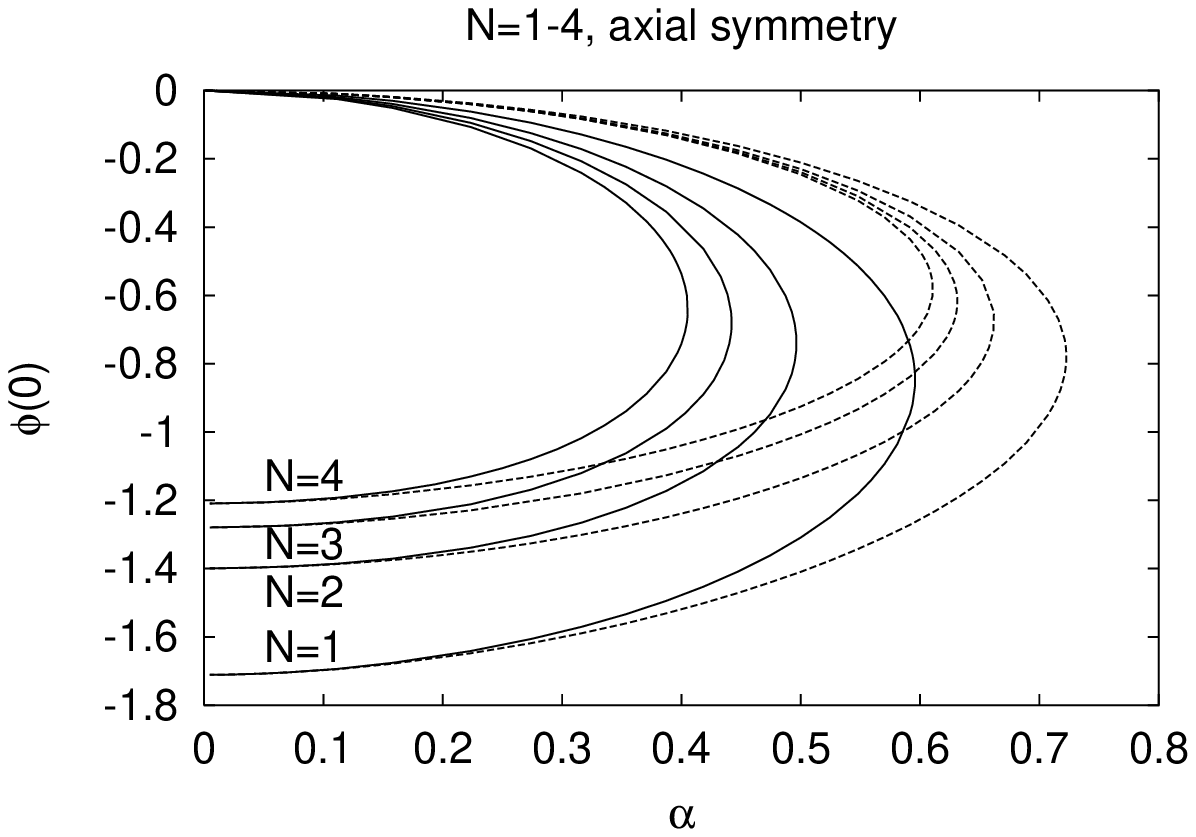} }
}\vspace{0.5cm} 
{\bf Fig.~1} \small
The dimensionless energy $E$ (a) and the 
value of the dimensionless dilaton field at the origin 
$\phi(0)$ (b)
of the spherically symmetric ($N=1$) and axially symmetric ($N=2-4$)
sphalerons
are shown as functions of the dilaton coupling constant $\alpha$ 
for the
Higgs masses $M_{\rm H}=M_{\rm W}$ (solid) and $M_{\rm H}=0$ (dashed).
\vspace{0.5cm}
}

Considering the limit of vanishing $\alpha$ on the upper branch,
we note that the scaled energy $\alpha E$ and 
the value of the dilaton function at the origin
$\phi(0)$ approach finite values, 
equal to the energy $E$, respectively $\phi(0)$,
of the spherically symmetric YMD solution \cite{lav}.
Indeed, introducing the scaled variable $\hat{r} = r/\alpha$, 
the Higgs field $\hat{\Phi} = \Phi \alpha$ and the gauge field 
$\hat{V}_i = V_i \alpha$, we arrive at an equivalent system
of equations,
in which the limit of vanishing Higgs field $\hat{\Phi}$
corresponds to the limit of vanishing $\alpha$ on the upper branch
of the original system. 
We exhibit the scaled energy $\alpha E$
as a function of $\alpha$ in Fig.~2.
\footnote{
Since the direct integration of the energy density
for the platonic sphalerons with $N=3$
is rather inaccurate due to the large numerical error of the solution,
we use instead the relation $\alpha E= D/\alpha$
to determine the scaled energy.}

\parbox{\textwidth}{
\centerline{
(a)\mbox{\epsfysize=5.0cm \epsffile{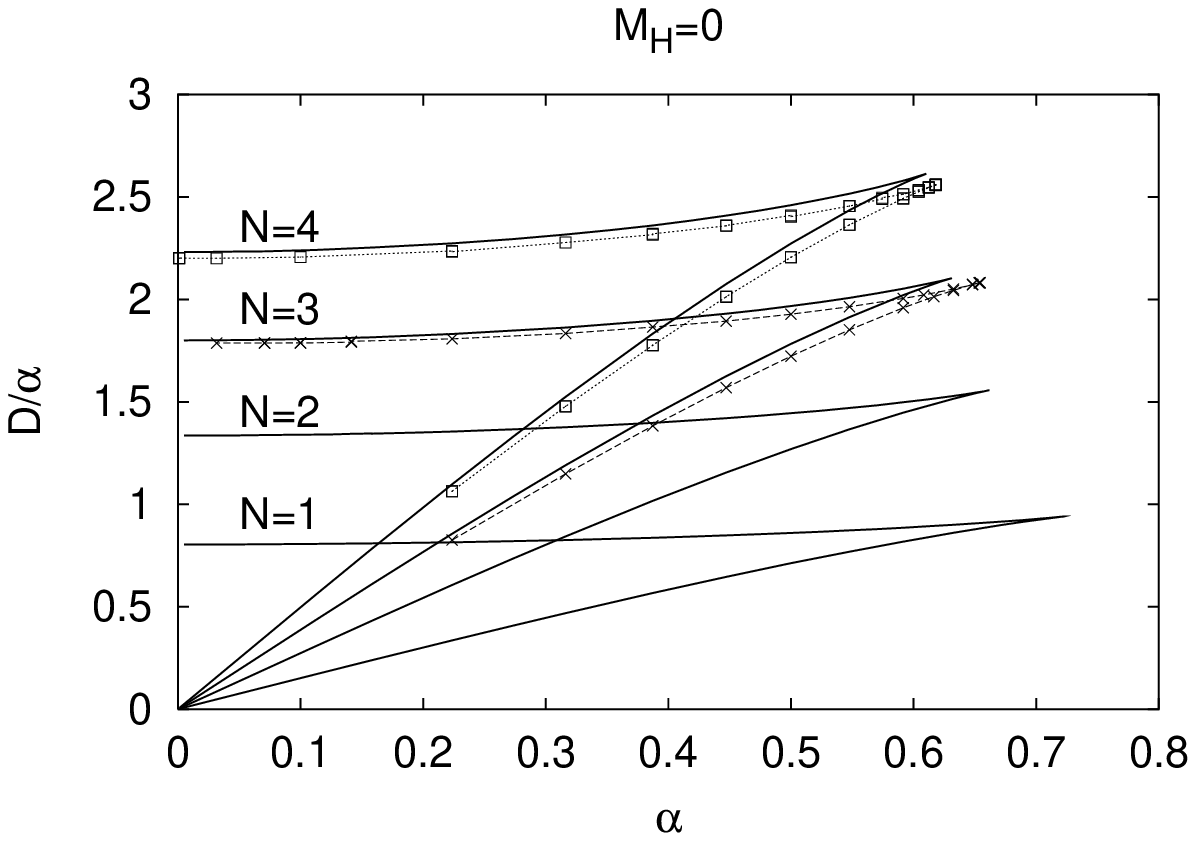} }
(b)\mbox{\epsfysize=5.0cm \epsffile{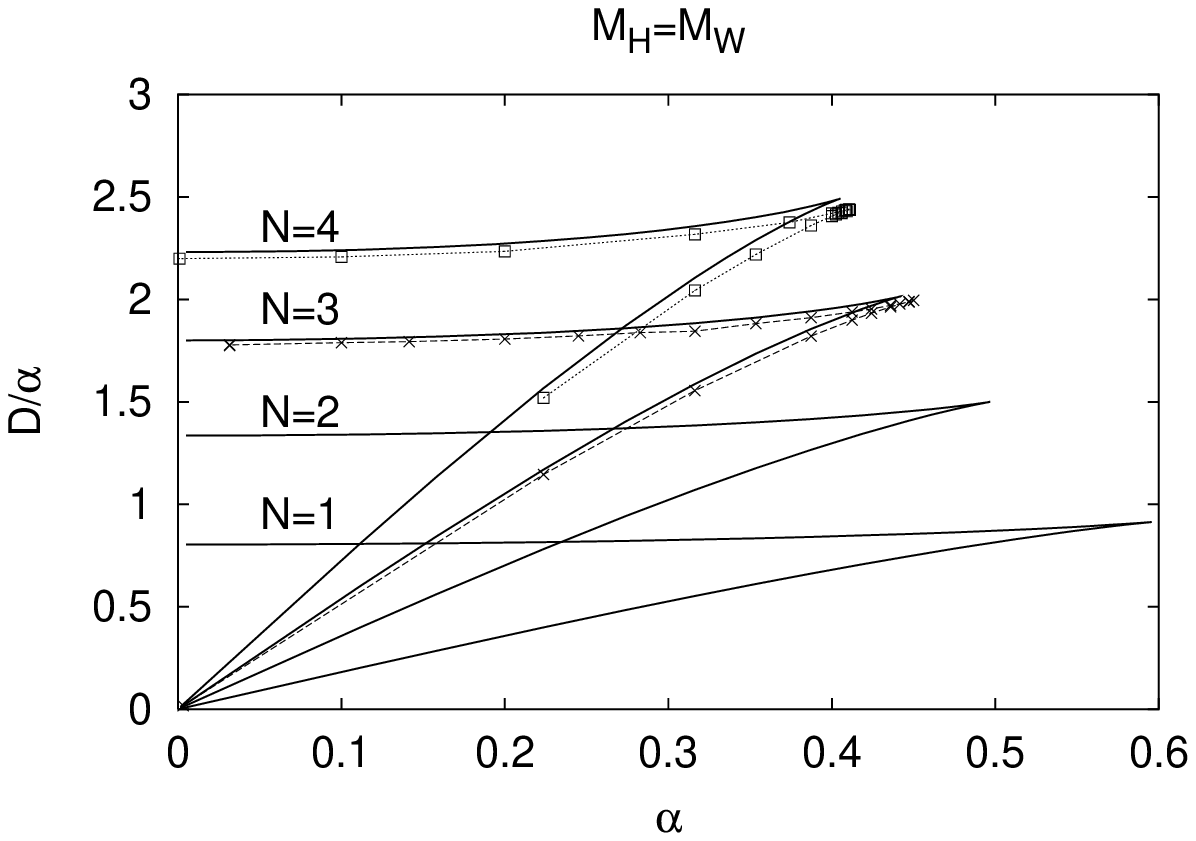} }
}\vspace{0.5cm}
{\bf Fig.~2} \small
The scaled energy $\alpha E$ 
of the spherically symmetric ($N=1$),
axially symmetric ($N=2-4$) and platonic sphalerons ($N=3$,4, represented
by symbols)
is shown as a function of $\alpha$ for the
Higgs masses $M_{\rm H}=0$ (a) and $M_{\rm H}=M_{\rm W}$ (b).
\vspace{0.5cm}
}

\subsection{Axially symmetric sphalerons}

For the axially symmetric sphalerons we observe the same pattern
of solutions as for the spherically symmetric sphalerons.
For a given $N$,
the lower branch of dilatonic sphalerons emerges from the
corresponding sphaleron of the Weinberg-Salam theory \cite{kksph},
and bifurcates with the upper branch at
a maximal value of the dilaton coupling constant $\alpha_{\rm max}$.
The maximal value $\alpha_{\rm max}$ decreases with increasing $N$ 
and with increasing $M_{\rm H}$.
As $\alpha$ tends to zero on the upper branch, the dilatonic sphalerons
tend to the axially symmetric YMD solutions \cite{kkYMD} after rescaling.
The energy and the value of the dilaton function at the origin 
of the axially symmetric sphaleron solutions with $N=2$, $3$, $4$
are shown in Fig.~1,
and their scaled energy is shown in Fig.~2.

\subsection{Platonic sphalerons}

To obtain better numerical accuracy for the platonic sphalerons, 
we make use of the discrete symmetries of the solutions,
to restrict the region of the numerical integration.
Let us first consider the tetrahedral $N=3$ solution. The rational map 
Eq.~(\ref{map1}) with $a=i$ leads to the unit vector 
\begin{equation}
\vec {n}_R= \left( -\frac{x^3+\sqt r y z}{r^3+\sqt x y z} \ , 
                   -\frac{y^3+\sqt r x z}{r^3+\sqt x y z} \ ,
                   -\frac{z^3+\sqt r x y}{r^3+\sqt x y z} 
	   \right) \ .
\end{equation}
To exploit the reflection symmetries about the $xz$- and $yz$-planes,
we introduce new coordinates $x'=(x+y)/\sqrt{2}$, $y'=(x-y)/\sqrt{2}$ 
and define a new unit vector 
\begin{eqnarray}
\vec {n}_{R'} & = & ( (n_R^1+n_R^2)/\sqrt{2},(n_R^1-n_R^2)/\sqrt{2}, n_R^3 ) 
\nonumber \\
              & = & \left(-x'\frac{{x'}^2 + 3 {y'}^2 + 2\sqt r z}
                                  {2r^3 +\sqt({x'}^2-{y'}^2)} \ , 
	                  -y'\frac{-({y'}^2 + 3 {x'}^2) + 2\sqt r z}
                                  {2r^3 +\sqt({x'}^2-{y'}^2)} \ , 
			   - \frac{ 2 z^3 + \sqt({x'}^2-{y'}^2)}
			          { 2 r^3 +  \sqt({x'}^2-{y'}^2)}
	   \right) \ .
\label{map1a}
\end{eqnarray}
Clearly, $n_{R'}^1$ is odd in $x'$ and even in $y'$, 
while $n_{R'}^2$ is even in $x'$ and odd in $y'$, 
and $n_{R'}^3$ is even in $x'$ and $y'$.
We then rename $x'=x$ and $y'=y$ and redefine 
$\tau_{R} := (\vec {n}_{R'})\cdot {\vec \tau}$.

Next let us turn to the cubic $N=4$ solution. 
The rational map Eq.~(\ref{map2}) leads to the unit vector
\begin{equation}
\vec {n}_R= \left(-\frac{(r^2-z^2)^2-2 z^2 r^2+2 x^2 y^2}
                        {{\cal N}_4} \ , 
                   \frac{\sqt (r^2+z^2)(x^2-y^2)}
                        {{\cal N}_4} \ , 
                   \frac{4 \sqt r x y z}
                        {{\cal N}_4} 
             \right) \ ,	 
\end{equation}
where ${\cal N}_4 =2(x^4+x^2 y^2+x^2 z^2+y^4+y^2 z^2+z^4)$.
In this case $n_R^1$ and  $n_R^2$  are even, while $n_R^3$ is odd 
in $x$, $y$, $z$.

We now suppose that the Higgs field $\Phi$ and the
gauge field $V_i$ possess the same reflection symmetries as 
$\tau_{R}$ and $\frac{i}{g} [\partial_i \tau_{R} ,\tau_{R}]$, respectively.
For $N=3$ it is then sufficient to solve the field equations for $x\ge 0$ and 
$y\ge 0$ only, while for $N=4$ we can restrict to 
$x\ge 0$, $y\ge 0$ and $z\ge 0$.

The boundary conditions at infinity must then be supplemented
by conditions at the other boundaries of the integration region.
At the origin $\bar{r}=0$, the gauge field and Higgs field components
have to vanish, and for the dilaton field we impose the conditon
$\partial_{\bar{r}} \phi = 0$.
The boundary conditions in the $xz$-plane ($\vphi=0$),  
the $yz$-plane ($\vphi=\pi/2$), on the $z$-axis 
($\theta=0\, , \pi$) and in the $xy$-plane ($\theta=\pi/2$, for $N=4$ only)
follow from the reflection symmetries of the functions. 
They are given in Table 1 for the gauge field and Higgs field components.
For the dilaton function the normal derivative has to vanish in 
the $xy$, $xz$ and $yz$-plane, and on the $z$-axis $\partial_\theta \phi =0$. 

\begin{table}
\begin{tabular}{|c|c|c|}
\hline
$N$ & $\vphi=0$ & $\vphi=\pi/2$
\\ \hline
 3 & \begin{tabular}{ccc}
      $\partial_\vphi \Phi_1=0$\ , & $\Phi_2$=0 \ , & $\partial_\vphi \Phi_3=0$ \ , \\
      $V_x^1=0$ \ , &  $\partial_\vphi V_x^2=0$  \ , &   $V_x^3=0$ \ ,\\
      $\partial_\vphi V_y^1=0$ \ , & $V_y^2=0$  \ , & $\partial_\vphi  V_y^3=0$ \ , \\     
      $V_z^1=0$ \ ,  &  $\partial_\vphi V_z^2=0$  &  \ ,  $V_z^3=0$ \ ,      
     \end{tabular} 
   & \begin{tabular}{ccc}
     $\Phi_1=0$\ , & $\partial_\vphi \Phi_2=0$\ ,  & $\partial_\vphi \Phi_3=0$\ ,  \\
      $V_x^1=0$  \ , &  $\partial_\vphi V_x^2=0$  \ ,  & $\partial_\vphi V_x^3=0$  \ ,  \\
      $\partial_\vphi V_y^1=0$  \ ,  &  $V_y^2=0$  \ ,  &  $V_y^3=0$  \ ,  \\ 
      $\partial_\vphi V_z^1=0$  \ , & $ V_z^2=0$   \ ,  & $  V_z^3=0$   \ , 
     \end{tabular} 
\\ \hline
4 & \begin{tabular}{ccc}
       $\partial_\vphi \Phi_1=0$ \ , &  $\partial_\vphi \Phi_2=0$ \ ,  &  $\Phi_3=0$ \ ,\\
       $V_x^1=0$ \ , &   $V_x^2=0$ \ ,  &  $\partial_\vphi V_x^3=0$ \ , \\
       $\partial_\vphi V_y^1=0$ \ , &   $\partial_\vphi V_y^2=0$ \ ,  &   $V_y^3=0$ \ , \\
       $V_z^1=0$ \ ,  &  $V_z^2=0$ \ , &   $\partial_\vphi V_z^3=0$ \ , 
     \end{tabular} 
   & \begin{tabular}{ccc}
       $\partial_\vphi \Phi_1=0$ \ ,  &  $\partial_\vphi \Phi_2=0$ \ ,  &   $\Phi_3=0$ \ , \\
       $\partial_\vphi V_x^1=0$   &  \ , $\partial_\vphi V_x^2=0 $\ ,   &   $ V_x^3=0 $\ , \\
       $V_y^1=0$ \ ,   &   $V_y^2=0$ \ ,   &  $\partial_\vphi  V_y^3=0$ \ , \\
       $V_z^1=0$ \ ,   &   $V_z^2=0$ \ ,   &  $\partial_\vphi  V_z^3=0$ \ .
     \end{tabular} 
 \\ \hline
    & $z$-axis & $xy$-plane \\
 \hline  
 3   & \begin{tabular}{ccc} 
         $\Phi_1=0$ \ , & $\Phi_2=0 $  \ , & $\partial_\theta \Phi_3=0$ \ ,\\
	 $V_x^1=0$ \ , & $\partial_\theta V_x^2=0$ \ , & $V_x^3=0$ \ , \\
	 $\partial_\theta V_y^1=0$ \ , & $V_y^2=0$ \ , & $V_y^3=0$ \ , \\
	 $V_z^1=0$ \ , & $V_z^2=0$ \ , & $V_z^3=0$ \ , 
     \end{tabular} 
     & 
     --
\\ \hline
4 & \begin{tabular}{ccc}
     $\partial_\theta \Phi_1=0$ \ , & $\partial_\theta \Phi_2=0$ \ ,& $\Phi_3=0$ \ , \\
     $V_x^1=0$ \ , & $V_x^2=0$ \ , & $V_x^3=0$ \ , \\
     $V_y^1=0$ \ , & $V_y^2=0$ \ , & $V_y^3=0$ \ , \\
     $V_z^1=0$ \ , & $V_z^2=0$ \ , & $\partial_\theta V_z^3=0$ \ , 
     \end{tabular} 
  & \begin{tabular}{ccc}
     $\partial_\theta \Phi_1=0$ \ , & $\partial_\theta \Phi_2=0$ \ ,& $\Phi_3=0$ \ , \\
     $V_x^1=0$ \ , & $V_x^2=0$ \ , & $\partial_\theta V_x^3=0$ \ , \\
     $V_y^1=0$ \ , & $V_y^2=0$ \ , & $\partial_\theta V_y^3=0$ \ , \\
     $\partial_\theta V_z^1=0$ \ , & $\partial_\theta V_z^2=0$ \ , & $V_z^3=0$ \ ,
     \end{tabular} 
\\ \hline
\end{tabular} 
\caption{ \small
The boundary conditions 
in the $xy-$, $xz-$, $yz$-plane, and on the $z$-axis
are given for the gauge and Higgs field components
of the platonic sphalerons with $N=3$ and $N=4$.}
\end{table}

We now turn to the numerical results.
For the platonic sphalerons,
we again observe the same pattern seen for the spherically
and axially symmetric sphalerons.
There are two branches of solutions which bifurcate at the 
maximal value $\alpha_{\rm max}$ of the coupling parameter $\alpha$,
where $\alpha_{\rm max}$ depends on $N$ and on the Higgs mass.
The branch with lower energy emerges from the platonic sphalerons 
of the Weinberg-Salam theory \cite{kkm}, 
while the dilatonic sphalerons on the upper branch
tend (after scaling) to YMD solutions with platonic symmetry \cite{kkmnew}.
The scaled energy of these solutions is also exhibited in Fig.~2,
and the value of their dilaton function at the origin
is shown in Fig.~3.

\parbox{\textwidth}{
\centerline{
(a)\mbox{\epsfysize=5.0cm \epsffile{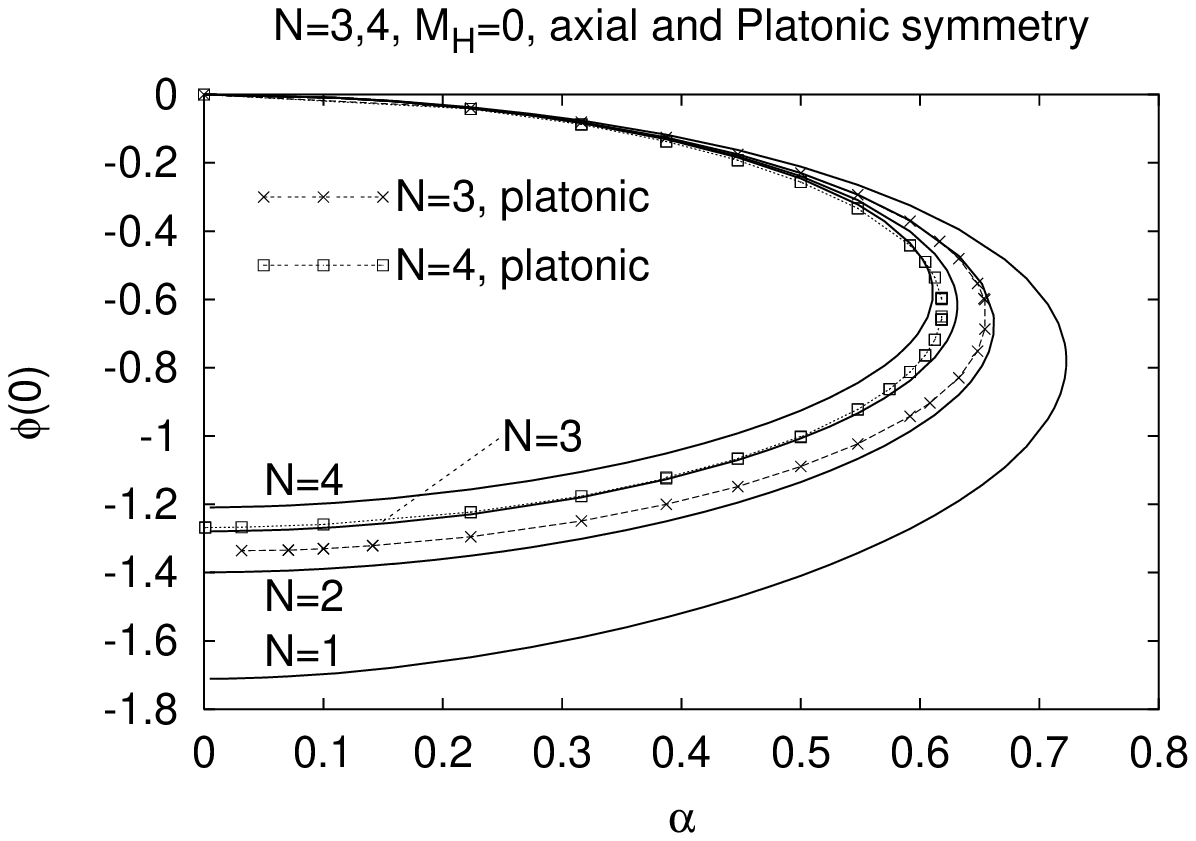} }
(b)\mbox{\epsfysize=5.0cm \epsffile{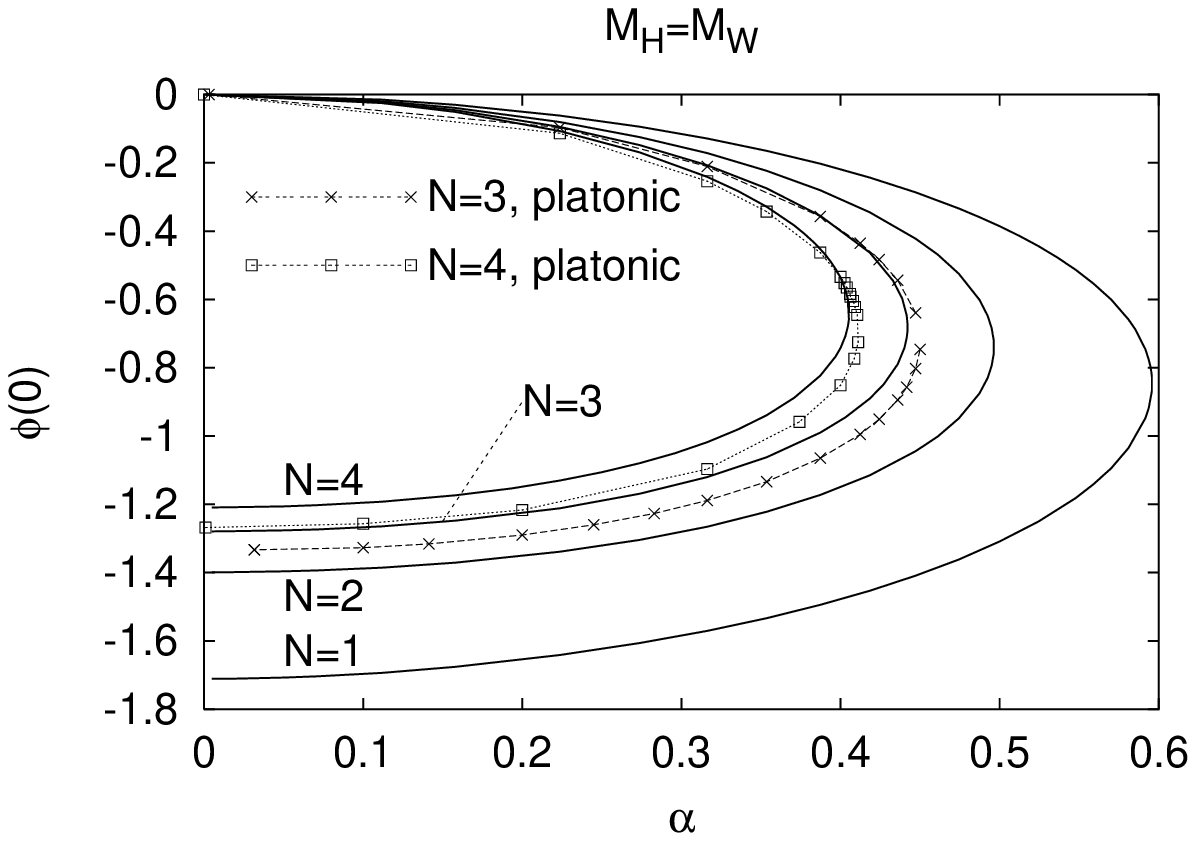} }
}\vspace{0.5cm}
{\bf Fig.~3} \small
The value of the dimensionless dilaton field at the origin $\phi(0)$ is
shown as a function of $\alpha$ for the platonic sphalerons 
($N=3, 4$), as well as for the spherically ($N=1$)
and axially symmetric sphalerons ($N=2-4$) 
for the Higgs masses $M_{\rm H}=0$ (a) and $M_{\rm H}=M_{\rm W}$ (b).
\vspace{0.5cm}
}

Comparison of the platonic sphalerons and the axially symmetric 
sphalerons reveals, that the platonic sphalerons exist 
for slightly larger values of the dilaton coupling constant $\alpha$
than the axially symmetric sphalerons with the same $N$.
For all values of $\alpha$, for which both types of
sphalerons coexist, the energy of the platonic sphalerons 
is of a similar magnitude, but slightly smaller
than the energy of the 
axially symmetric sphalerons with the same $N$.

Interestingly, as seen in Fig.~3,
the values of the dilaton function at the origin $\phi(0)$
of the $N=3$ platonic sphaleron solutions agree 
well with those of the $N=2$ and $N=3$ axially symmetric solutions
on the lower branch for $M_{\rm H}=0$ and $M_{\rm H}=M_{\rm W}$,
respectively,
and those of the $N=4$ platonic sphaleron solutions agree
well with those of the $N=4$ axially symmetric solutions
on the lower branch and 
with those of the $N=3$ axially symmetric solutions
on the upper branch.

Defining the dimensionless energy density $\varepsilon$ 
of the sphalerons by
\begin{equation}
E= \int \varepsilon (\vec x) dx dy dz
\ , \end{equation}
we present surfaces of constant energy density $\varepsilon$ in Fig.~4
for the $N=3$ sphaleron with $M_{\rm H}=M_{\rm W}$,
and for the $N=4$ sphaleron with $M_{\rm H}=0$,
where the values of $\alpha$
are chosen close to the respective maximal value $\alpha_{\rm max}$.
Clearly, the energy density of these solutions
exhibits tetrahedral, and cubic symmetry, respectively.
As observed previously \cite{kkm}, the shape of the energy density 
is determined primarily by the rational map. 

\parbox{\textwidth}{
\centerline{
(a)\mbox{\epsfysize=8.0cm \epsffile{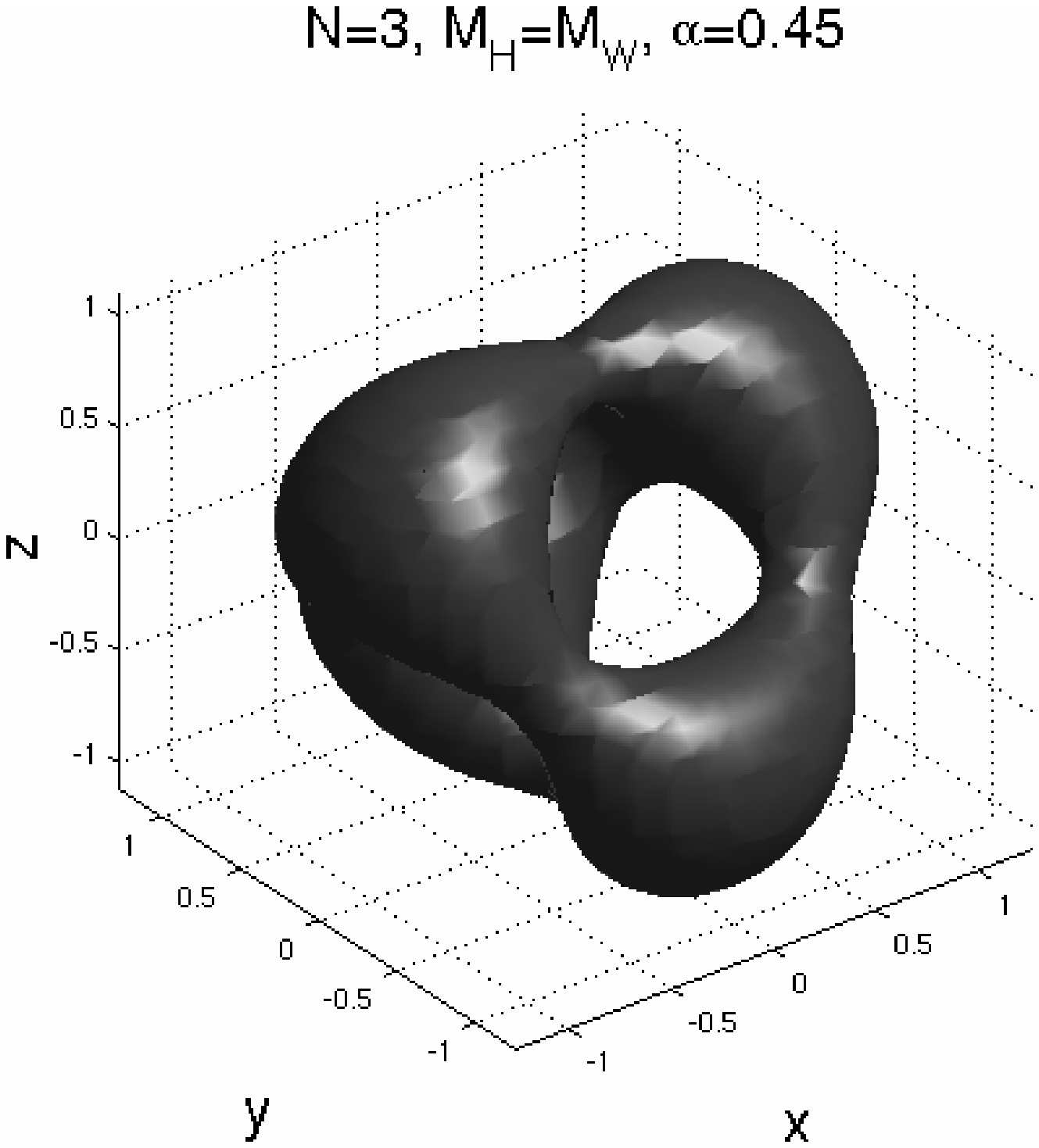} }
(b)\mbox{\epsfysize=8.0cm \epsffile{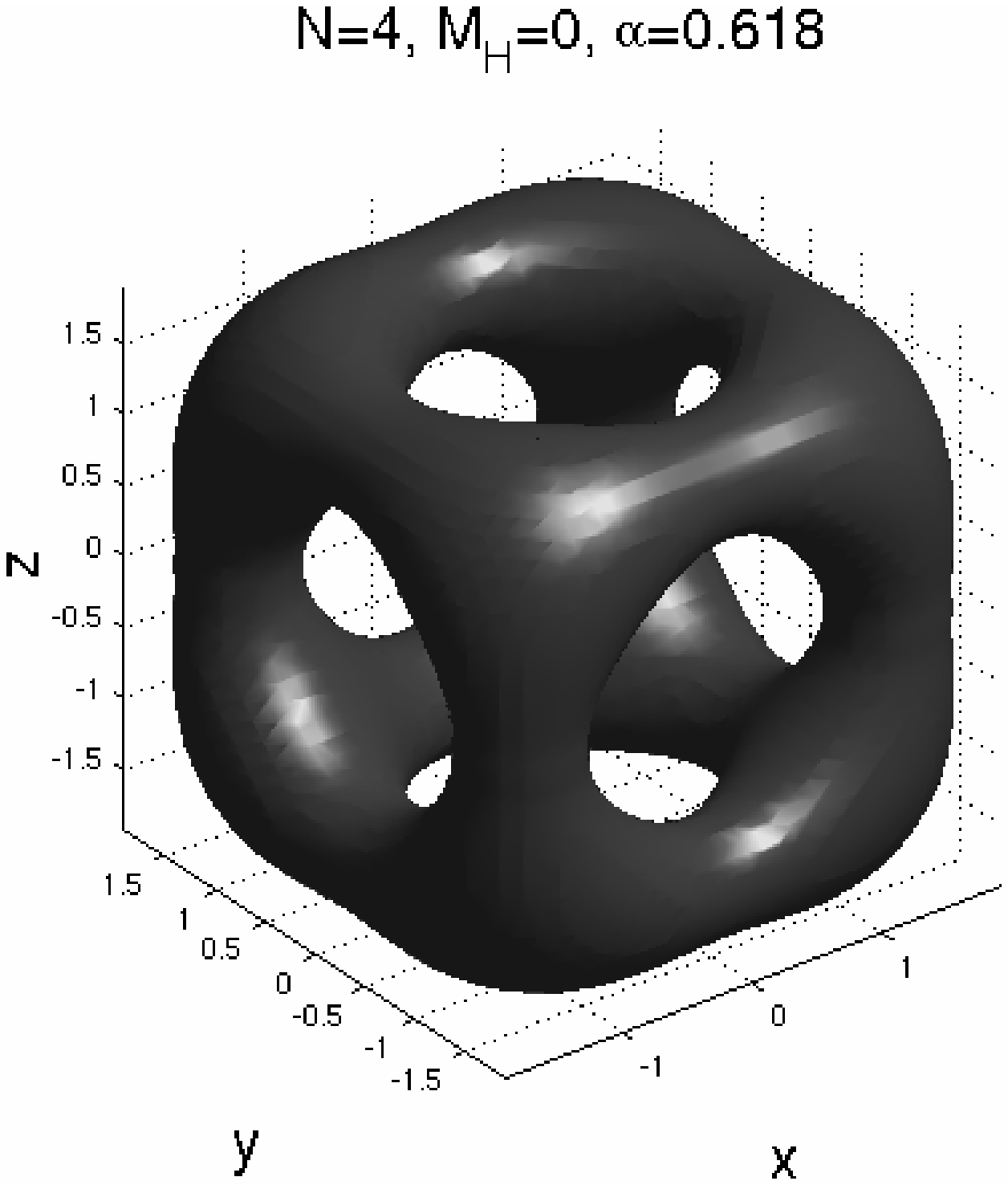} }
}\vspace{0.5cm}
{\bf Fig.~4} \small
The dimensionless energy density $\epsilon$
of the platonic sphalerons 
with $N=3$ and $\alpha=0.45$ (a) 
and $N=4$ and $\alpha=0.618$ (b)
is shown for the Higgs mass $M_{\rm H}=M_{\rm W}$,
close to the respective maximal value $\alpha_{\rm max}$.
\vspace{0.5cm}
}

For the tetrahedral sphaleron the modulus of the Higgs field has five nodes,
four located on the diagonals close to the maxima of the energy density,
and one located at the origin \cite{kkm}. The distance of the nodes
on the diagonals from the origin is exhibited in
Fig.~5 as a function of $\alpha$.
The cubic sphaleron, in contrast, has a single node located at the origin
\cite{kkm}.

\parbox{\textwidth}{
\centerline{
\mbox{\epsfysize=5.0cm \epsffile{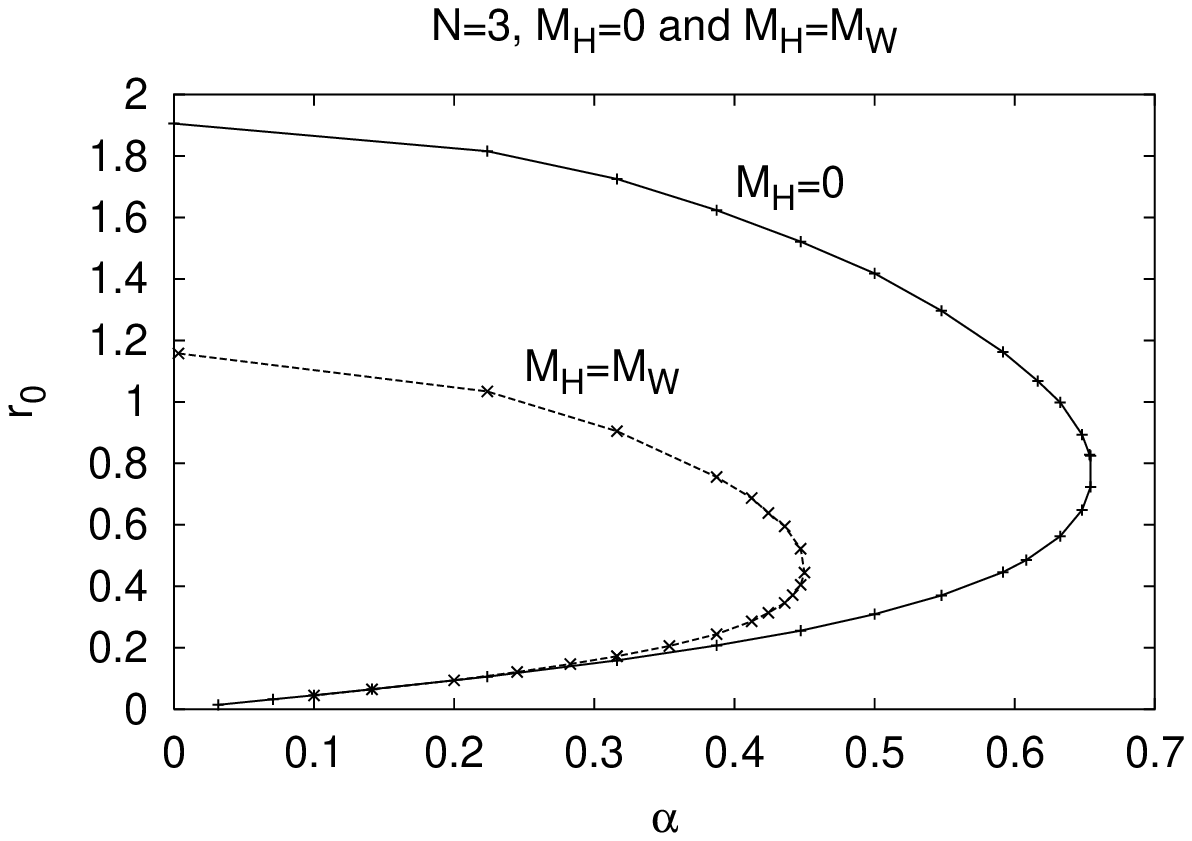} }  
}\vspace{0.5cm}
{\bf Fig.~5} \small
The distance of the node of the 
modulus of the Higgs field on the diagonal from the origin
is shown as a function of $\alpha$ for tetrahedral sphalerons ($N=3$) and
Higgs masses $M_{\rm H}=0$ (solid) and $M_{\rm H}=M_{\rm W}$ (dashed).
\vspace{0.5cm}
}

We give a brief summary of the main results in Table 2,
where the scaled energy and the energy are shown
at the limiting value
$\alpha=0$ on both branches of solutions, respectively,
as well as the maximum value of the
dilaton coupling constant $\alpha_{\rm max}$
for all solutions considered.

\begin{table}
\begin{center}
\begin{tabular}{|c|c|c|c|c|c|} \hline\hline
    &  & \multicolumn{2}{c|}{$M_{\rm H}=0$ }
       & \multicolumn{2}{c|}{$M_{\rm H}=M_{\rm W}$}  \\
\hline
$N$ &  $\alpha E_0$ & $\alpha_{\rm max}$ & $E_0$ &
                      $\alpha_{\rm max}$ & $E_0$ \\  
\hline
$1$   &  $0.80$   & $0.722$ & $1.52$ & $0.596$ & $1.82$\\   
$2^*$ &  $1.33$   & $0.662$ & $2.75$ & $0.496$ & $3.60$\\   
$3  $ &  $1.80$   & $0.654$ & $3.91$ & $0.447$ & $5.33$\\   
$3^*$ &  $1.80$   & $0.631$ & $3.89$ & $0.442$ & $5.44$\\   
$4  $ &  $2.20$   & $0.618$ & $4.84$ & $0.411$ & $6.80$\\  
$4^*$ &  $2.23$   & $0.601$ & $4.99$ & $0.405$ & $7.34$\\  
\hline
\hline
\end{tabular}
\caption{ \small
The scaled energy $\alpha E_0$ on the upper branch
of the limiting $\alpha=0$ YMD solution,
and the maximal value
of the dilaton coupling constant $\alpha_{\rm max}$
and the energy $ E_0$ on the lower branch
of the limiting $\alpha=0$ YMH solution
are shown for the spherically symmetric ($N=1$),
axially symmetric ($N=2-4$)
and platonic sphalerons ($N=3$,4)
for the
Higgs masses $M_{\rm H}=0$ and $M_{\rm H}=M_{\rm W}$.
$N^*$ configurations represent axially symmetric sphalerons.
}
\end{center}
\end{table}

\section{Conclusions}

We have constructed numerically sphaleron solutions of YMHD theory,
possessing spherical, axial, tetrahedral and cubic symmetry.
For all these sphalerons two branches of solutions exist,
thus they reveal the same general dependence on the
dilaton coupling constant.
When the dilaton coupling constant $\alpha$ is increased from zero,
the lower branch of dilatonic sphalerons emerges
from the corresponding sphaleron of Weinberg-Salam theory.
The lower branch extends up to a maximal value $\alpha_{\rm max}$  
of the dilaton coupling constant,
where it bifurcates with the upper branch of sphaleron solutions.
The upper branch then extends backwards to $\alpha=0$,
where it ends in a sphalerons solution of YMD theory (after rescaling).
While spherically and axially symmetric YMD solutions were known before
\cite{lav,kkYMD}, we have here obtained first evidence for the existence 
of YMD solutions with platonic symmetries \cite{kkmnew}.

Viewing this study as a first step towards obtaining extended gravitating
solutions without rotational symmetries, let us reconsider the
simpler spherically symmetric case,
since gravitating spherically symmetric sphalerons
have been studied before \cite{greene,vg,yves}. 
These exhibit the same general coupling constant dependence,
as the dilatonic sphalerons studied here.
Even the maximal values of the coupling constants
for gravity and the dilaton are close,
as observed before for monopoles \cite{gmono,forgacs}.
For the dilaton coupling constant we find
$\alpha_{\rm max}=0.722$ and $0.596$
for $M_{\rm H}=0$ and $M_{\rm H}=M_{\rm W}$, respectively,
while the corresponding gravitational coupling constants
are $\alpha_{\rm max}=0.750$ \cite{vg}
and $\alpha_{\rm max}=0.619$ \cite{yves} \footnote{
Note, that refs.~\cite{vg,yves} employ different definitions for
the coupling constants.}.
%vg: 0.281: factor of 2 and then square root
%yves: 0.3095: factor of 2
We thus expect, that gravitating platonic sphalerons
will exhibit the same main features as the dilatonic
sphalerons with discrete symmetries, studied here.

{\bf Acknowledgement}: 

B.K.~gratefully acknowledges support by the DFG under contract
KU612/9-1, and K.M.~by the Research Council of Norway under 
contract 153589/432.

%\vfill\eject

\end{document}